\documentclass[12pt]{article}
         
\usepackage{amsmath,amssymb} 
\usepackage{pictfile}     
\usepackage[mathscr]{eucal}       
\usepackage{latexsym,amsfonts,amstext,verbatim}            
\usepackage[dvips]{graphicx}

\numberwithin{equation}{section}

\newcommand{\gpr}{g^{\prime}}
\newcommand{\Dpr}{D^{\prime}}

\newcommand{\upr}{u^{\prime}}

\newcommand{\omegapr}{\omega^{\prime}}
\newcommand{\tg}{\tilde{g}}

\newcommand{\tD}{\tilde{D}}

\newcommand{\tG}{\tilde{G}}

\newcommand{\sfA}{\sf A}
\newcommand{\sfB}{\sf B}
\newcommand{\sfC}{\sf C}
\newcommand{\sfD}{\sf D}
\newcommand{\sfE}{\sf E}
\newcommand{\sfF}{\sf F}
\newcommand{\sfG}{\sf G}

\newcommand{\bsm}{\boldsymbol}

\newcommand{\rmi}{\mathrm{i}}
\newcommand{\rme}{\mathrm{e}}
\newcommand{\rmd}{\mathrm{d}}
\newcommand{\bR}{\mathbb{R}}

\newcommand{\frH}{\mathfrak{H}}

\newcommand{\bma}{\bsm{a}}
\newcommand{\bmr}{\bsm{r}}
\newcommand{\bmp}{\bsm{p}}
\newcommand{\bmv}{\bsm{v}}

\newcommand{\bmP}{\bsm{P}}

\newcommand{\bmS}{\bsm{S}}
\newcommand{\bmJ}{\bsm{J}}
\newcommand{\bmK}{\bsm{K}}

\newtheorem{assump}{Assumption}

\newtheorem{prop}{Proposition}

\renewcommand{\[}{\hspace{0.0125em}\relax}

\begin{document}

 \title{\large\bf Little-used Mathematical Structures\\[0.7mm]in Quantum
Mechanics\\[2mm] I. Galilei Invariance and the \emph{welcher Weg} Problem}

\author{ {R N Sen}\\
{\normalsize Department of Mathematics}\\[-0.7mm]
{\normalsize Ben-Gurion University of the Negev}\\[-0.7mm] 
{\normalsize 84105 Beer Sheva, Israel}\\
{\normalsize\tt E-mail: rsen@cs.bgu.ac.il}\\[1mm]
{\normalsize 21 January 2012}}
\date{}

\maketitle
\thispagestyle{empty}

\vspace{-5mm}
\begin{abstract}

\vspace{2mm}
\noindent
Results of the \emph{welcher Weg} experiment of D\"urr, Nonn and
Rempe are explained by using ray representations of the Galilei
group.  The key idea is that the state of the incoming atom be
regarded as belonging to an irreducible unitary ray representation of
this group.  If this is the case, interaction with an interferometer
with a which-way detector must split this state into the direct sum
of two states belonging to representations with different internal
energies.  (While the zero of internal energy is arbitrary, the
difference between two internal energies is well-defined and is invariant
under unitary transformations.)  The state of the outgoing atom will
then be a superposition of two mutually orthogonal states, so that
there will be no interference. Neither complementarity nor entanglement
plays a role in this explanation.  Furthermore, in atom
interferometry it is not enough for a quantum eraser to erase the
internal energy difference; to restore interference, two copies of a
representation have to be collapsed into one. In a direct sum of
copies of the same representation, copies of the same state will
still be orthogonal.  These assertions may be testable, and two new
atom interferometry experiments are suggested.  One of them is an
`own-goal' experiment which may decisively refute the explanation
offered here, and restore the aura of mystery that this paper tries
to dispel.

\end{abstract}
\pagebreak

This is the first of three papers on mathematical structures in
quantum mechanics that have been known for four to six
decades,\footnote{By quantum mechanics we mean von Neumann's Hilbert
space formulation of it. The conceptual subtleties that underlie the
accompanying paper cannot be expressed in Dirac's formulation.} but
have not been fully exploited by physicists. These structures reveal
new possibilities for the interpretation of experiments that have
been performed, and suggest new experiments that may shed light on
unresolved or disputed problems in the foundations of quantum
mechanics. The present paper re-examines the \emph{welcher Weg}
problem from the viewpoint of Galilei invariance. The accompanying
paper examines physical implications of the existence of inequivalent
(irreducible unitary) representations of the canonical commutation
relations for a \emph{finite} number of degrees of freedom. The last
paper will investigate consequences of the existence of dense sets of
analytic functions on some concrete Hilbert spaces used in physics.
The existence of such sets was first noted by Stone in the late 1920s
or early 1930s; Hilbert spaces of analytic functions were explicitly
constructed by Bargmann in 1961.

\section{The \emph{welcher Weg} problem}

The welcher Weg problem appears to have been suggested by the
\emph{Feynman Lectures on Physics}.  In Chapter 1 of Part III of
these Lectures \cite{FLS2006}, Feynman prepares his undergraduate
audience for the shock of quantum mechanics by discussing two
double-slit gedankenexperiments with electrons. The first experiment
is the standard one; an interference pattern is gradually built up as
electrons strike the detection screen.  In the second experiment,
`which way' an electron takes is determined by shining a light beam
on it at the slits. The pattern that builds up now no longer shows
interference.  The loss of interference is explained by the
uncertainty principle. Feynman goes on to affirm that ``No one has
ever found (or even thought of) a way around the uncertainty
principle''. 

However, in 1991 Scully, Englert and Walther (hereafter SEW) claimed
that ``\ldots we have found a way around this position-momentum
uncertainty obstacle\ldots That is, we have found a way\ldots to
obtain which-path or particle-like information without scattering or
otherwise introducing large uncontrolled phase factors into the
interfering beams.''  They went on to state that, in their
gedankenexperiment, ``The principle of complementarity is manifest
although the position-momen\-tum uncertainty relation plays no role''
\cite{SEW1991}. 

The way around that SEW had found consisted of using cold atoms as
particles in a one-particle interference experiment. An excited atom
(travelling at a low speed) may be induced to emit a photon in a
resonant cavity in its path. If the parameters are right, emission of
the photon will have a negligible effect on the atom's linear
momentum, and therefore on its de Broglie wavelength; the effect on
the interference pattern should be negligible.  At the same time, the
emitted photon will reveal which path the atom took.

In 1998, D\"urr, Nonn and Rempe (hereafter DNR) performed the actual
experiment \cite{DNR1998} (now regarded as definitive) which, as we
shall see, differed from the one suggested by SEW in some essential
aspects.  Their findings agreed with the expectations of SEW; the
mere possibility of detection of the path taken by an atom resulted
in the loss of interference, and the effect could not be explained by
an appeal to the position-momentum uncertainty relation. The title
that DNR gave to their paper was ``Origin of quantum-mechanical
complementarity probed by a `which-way' experiment in an atom
interferometer''.

The chief aim of the present paper is to provide an explanation of
the phenomena observed by DNR that is based on the theory of ray
representations of the inhomogeneous Galilei group.  This theory can
describe atoms in different energy levels, but there is an unexpected
subtlety. The explanation, which also predicts phenomena that cannot
be accounted for by the DNR model, is quite independent of any notion
of complementarity. Additionally, two new experiments are suggested;
one of them, fittingly called the \emph{own-goal experiment}, may
refute the proposed explanation quite decisively.  The other may
discriminate between the DNR explanation and the one offered here.

The plan of the work is as follows.\footnote{There have been quite a
number of works on the subject before SEW, between SEW and DNR, and
after DNR. We have chosen to focus on DNR and SEW for reasons
mentioned above. References to earlier works will be found in SEW and
DNR. References to some later works may be found in a 2010 article by
Ferrari and Braunecker \cite{FB2010}.} In {Sec.}~\ref{SEC-SEW} we
review, briefly, the gedankenexperiment of SEW.  {In
Sec.}~\ref{SEC-DNR} we review the experiment of DNR.  This is
followed by {Sec.}~\ref{SEC-MAIN}, the main section of this paper.
In it we use the theory of ray representations of the Galilei group
to explain the experimental results of DNR without using any notion
of complementarity. In {Sec.}~\ref{SEC-DISCUSS}, we compare our
explanation with that of DNR, which is modelled on a moving two-level
atom.  The new experiments are suggested in Sec.~\ref{SEC-EXPTS}, the
last section of the paper.  The first, discussed in
\ref{SUBSEC-OWN-GOAL}, is the own-goal experiment that may decisively
refute the explanation based on Galilei invariance.  The second is a
variant of the DNR experiment, and includes a quantum eraser which
may enable it to distinguish between the Gaillei invariance and the
DNR explanations. There are two Appendices. In the first, we collect
together the key definitions and formulae of the theory of ray
representations, and recapitulate the notion of unitary equivalence
for these representations. In the second, we provide some references
to (i)~quantum optics and the manipulation of particles by light, and
(ii)~complementarity and uncertainty, subjects which form the
experimental and theoretical backdrops to this paper.

\section{The gedankenexperiment of Scully, Englert and Walther}
\label{SEC-SEW}

\begin{figure}[ht]
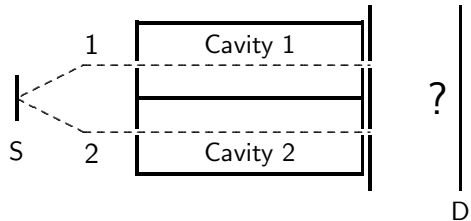


\footnotesize
\beginpicture
\small
\setcoordinatesystem units <1mm,1mm>

\setplotarea x from -55 to 45, y from -20 to 20

\linethickness=1pt

\putrule from 0 -10 to 30 -10
\putrule from 0 10 to 30 10
\putrule from 0 0 to 30 0 %

\putrule from 30 -10.3 to 30 -5  %
\putrule from 30 -4 to 30 4      %
\putrule from 30 5 to 30 10.3    %

\putrule from 31 -12.3 to 31 -5  %
\putrule from 31 -4 to 31 4      %
\putrule from 31 5 to 31 12.3    %

\putrule from 0 -10.3 to 0 -5   %
\putrule from 0 -4 to 0 4       %
\putrule from 0 5 to 0 10.3     %

\linethickness=1pt

\putrule from -16 -3 to -16 3

\putrule from 43 -12.3 to 43 12.3

\linethickness=0.5pt

\setdashes <.9mm>

\plot  -7 4.5  31 4.5 /
\plot  -7 -4.5 31 -4.5 /
\plot  -7 -4.5  -16 0  -7 4.5 /

\put {\footnotesize\sf D}      at 43 -15
\put {\footnotesize\sf S}     at -16 -7
\put {\footnotesize\sf Cavity 2} at 15 -7.5
\put {\footnotesize\sf Cavity 1} at 15 7
\put {\sf  2} at -6 -7.3
\put {\sf  1} at -6 7.2
\put {\Large\sf ?} at 40 0

\endpicture
\caption{Scheme of the SEW \emph{gedankenexperiment}}
\label{FIG-SEW}
\end{figure}

The experimental scheme of SEW is a double-slit interferometer
modified by the placement of a resonant cavity just before each slit,
shown schematically in Fig.\ \ref{FIG-SEW}. The two cavities, each
tuned to resonate at 21 GHz and prepared in the zero-photon state,
together constitute the which-way detector. The two paths are marked
{\sf 1} and {\sf 2} in the figure. Formulae (\ref{SCULLY-1}) and
(\ref{SCULLY-2}) are taken from the section on \emph{Gedanken
experiments illustrating complementarity} in \cite{SEW1991}. 

If the resonant cavities are not present, the state vector of the
atom, after it emerges from the double slit, will be will be given by
Eq.\ (4) of \cite{SEW1991}, namely (we adhere to their notation):
\begin{equation}\label{SCULLY-1}
\Psi(\bmr) = \displaystyle\frac1{\sqrt2}[\psi_1(\bmr)+\psi_2(\bmr)]
|i\rangle.
\end{equation}
In (\ref{SCULLY-1}), $\bmr$ is the coordinate of the centre of mass
and $|i\rangle$ the internal state of the atom. The subscripts $1$
and $2$ refer to paths 1 and 2 (Fig.\ \ref{FIG-SEW}).  When the
which-way detector is present, and the atom has emitted a photon in
one of the cavities, (\ref{SCULLY-1}) changes to (Eq.\ (6) in
\cite{SEW1991})

\begin{equation}\label{SCULLY-2}
\Psi(\bmr) = \displaystyle\frac1{\sqrt2}[\psi_1(\bmr)|1_10_2\rangle
+\psi_2(\bmr)|0_11_2\rangle]|b\rangle,
\end{equation}
where $|0_11_2\rangle$ is the state of the which-way detector with no
photon in cavity 1 and one photon in cavity 2, and similarly for
$|1_10_2\rangle$. The atom is prepared in the state $|a\rangle$ and
makes the transition $a \rightarrow b$ in the which-way detector.
The authors write that: ``Please note that unlike (4) this
$\Psi(\bmr)$ is not a product of two factors, one referring to the
atomic and the other to photonic degrees of freedom. The system and
the detector have become entangled by their interaction.''  The
orthogonality condition $\langle0_11_2|1_10_2\rangle = 0$ will now
cause the interference term in this $|\Psi(\bmr)|^2$ to vanish.

The validity of (\ref{SCULLY-2}) may be disputed, but we shall not
enter into this dispute, because (\ref{SCULLY-2}) is \emph{not} used
by DNR to interpret the results of the experiment they actually
performed. We now turn to this experiment.


\section{The experiment of D\"urr, Nonn and Rempe}
\label{SEC-DNR}

\begin{figure}[ht]
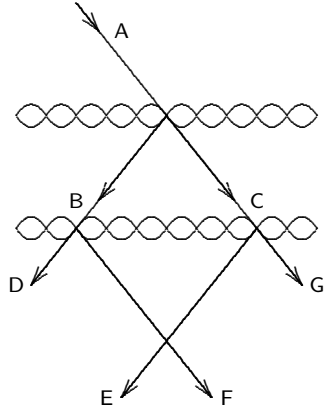


\beginpicture
\scriptsize

\setcoordinatesystem units <1mm,.75mm>

\setplotarea x from -70 to 30, y from -40 to 50

\linethickness=1pt

\setquadratic

\plot -20 20  -18 18  -16 20  -14 22  -12 20  -10 18  -8 20  -6 22  -4 20
-2 18  0 20  2 22  4 20  6 18  8 20  10 22  12 20  14 18  16 20  18 22  20 20 /

\plot -20  20  -18 22  -16 20  -14 18  -12 20  -10 22  -8 20  -6 18
-4 20   -2 22  0 20
2 18  4 20  6 22  8 20  10 18  12 20  14 22  16 20  18 18  20 20  /



\plot -20 0  -18 -2  -16 0  -14 2  -12 0  -10 -2  -8 0  -6 2  -4 0
-2 -2  0 0  2 2  4 0  6 -2  8 0  10 2  12 0  14 -2  16 0  18 2  20 0 /

\plot -20 0  -18 2  -16 0  -14 -2  -12 0  -10 2  -8 0  -6 -2  -4 0
-2 2  0 0  2 -2  4 0  6 2  8 0  10 -2  12 0  14 2  16 0  18 -2  20 0 /

\setlinear

\arrow <2.5mm> [0.25,0.6] from -12 40 to -9 35 

\arrow <2.5mm> [0.25,0.6] from 0 20 to 9 5 
\arrow <2.5mm> [0.25,0.6] from 0 20 to -9 5 

\arrow <2.5mm> [0.25,0.6] from 12 0 to -6 -30 
\arrow <2.5mm> [0.25,0.6] from 12 0 to 18 -10 

\arrow <2.5mm> [0.25,0.6] from -12 0 to -18 -10 
\arrow <2.5mm> [0.25,0.6] from -12 0 to 6 -30

\plot -12 40  18 -10 /
\plot 0 20  -18 -10 /


\plot 12 0  -6 -30 /
\plot -12 0  6 -30 /


\put {\sf A} at -6 35
\put {\sf B} at -12 5
\put {\sf C} at 12 5
\put {\sf D} at -20 -10
\put {\sf E} at -8 -30
\put {\sf F} at 8 -30
\put {\sf G} at 20 -10

\endpicture
\caption{The DNR experiment without which-way detection}
\label{FIG-DNR-1}
\end{figure}

\begin{figure}[ht]
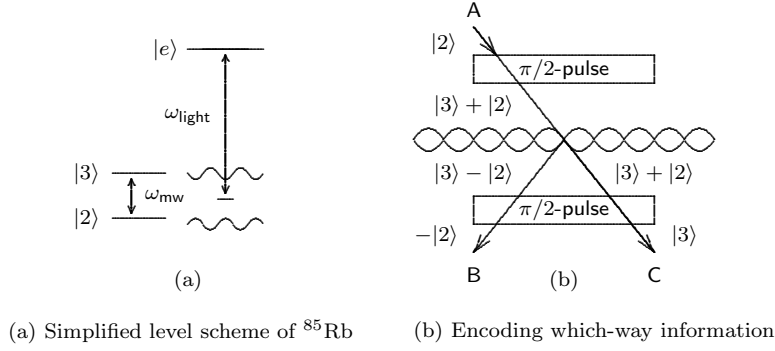


\beginpicture

\setcoordinatesystem units <1mm,.75mm>

\setplotarea x from -90 to 20, y from -20 to 50

\linethickness=1pt

\setquadratic

\plot -20 20  -18 18  -16 20  -14 22  -12 20  -10 18  -8 20  -6 22  -4 20
-2 18  0 20  2 22  4 20  6 18  8 20  10 22  12 20  14 18  16 20  18 22  20 20 /

\plot -20  20  -18 22  -16 20  -14 18  -12 20  -10 22  -8 20  -6 18
-4 20   -2 22  0 20
2 18  4 20  6 22  8 20  10 18  12 20  14 22  16 20  18 18  20 20  /



\setlinear

\arrow <2.5mm> [0.25,0.6] from -12 40 to -9 35 

\arrow <2.5mm> [0.25,0.6] from 0 20 to 12 0 
\arrow <2.5mm> [0.25,0.6] from 0 20 to -12 0 %

\plot -12 40  12 0 /

\plot -12 35  12 35  12 30  -12 30  -12 35 / 
\plot -12 10  12 10  12 5  -12 5  -12 10 /

\scriptsize
\put {$|2\rangle$} at -16 37  
\put {$-|2\rangle$} at -17 3   
\multiput {$|3\rangle + |2\rangle$} at -12 26  12 14 /
\put {$|3\rangle - |2\rangle$} at -12 14
\put {$|3\rangle$} at 16 3


\multiput {\sf $\pi/2$-pulse} at 0 7.5  0 32.5 /


\put {\sf A} at -12 43
\put {\sf B} at -12 -4
\put {\sf C} at  12 -4
\setquadratic

\plot   -50 14  -49 13  -48 14  -47 15  -46 14  -45 13  -44 14  -43 15
-42 14  -41 13  -40 14  /

\plot   -50 5  -49 6  -48 5  -47 4  -46 5  -45 6  -44 5  -43 4
-42 5  -41 6  -40 5  /

\setlinear

\plot -46 9.5  -44 9.5 /
\plot -50 36  -40 36 /

\arrow <1mm> [0.5,1] from -57.5 7 to -57.5 13
\arrow <1mm> [0.5,1] from -57.5 13 to -57.5 7

\arrow <1mm> [0.5,1] from -45 10.5 to -45 35 
\arrow <1mm> [0.5,1] from -45 35 to -45 10.5

\plot -60 14  -53 14 /
\plot -60 6   -53 6  /

\plot -48 36  -42 36 /

\scriptsize

\put {$|e\rangle$} at -53 36
\put {$|3\rangle$} at -63.5 14
\put {$|2\rangle$} at -63.5 6

\put{$\omega_{\sf light}$} at -50 24
\put{$\omega_{\sf mw}$} at -53 10

\put {\scriptsize (a)}  at -50 -5
\put {\scriptsize(b)}   at  0 -5

\put {\scriptsize (a) Simplified level scheme of $^{85}$Rb} at -51
-14 

\put {\scriptsize (b) Encoding which-way information} at 4 -14

\endpicture
\caption{The which-way detection scheme of DNR}
\label{FIG-DNR-2}
\end{figure}

In the experiment of D\"urr, Nonn and Rempe \cite{DNR1998}, a
monochromatic beam {\sfA} of $^{85}$Rb atoms is split into two, a
transmitted beam {\sfC} and a Bragg-refracted beam of the first order
{\sfB} using a standing light wave.\footnote{The idea of reflecting
electrons through a light crystal, which became feasible only with
intensities available with lasers, was thought of by Kapitza and
Dirac in 1933 \cite{KD1933}.  The reflection and refraction of matter
waves by standing light waves has become known as the
\emph{Kapitza-Dirac effect} \cite{B2000}. } The reflectivity of the
`light crystal' is determined by the intensity of the standing wave,
and is adjusted to $\sim50\%$.  Beams {\sfB} and {\sfC} are again
split into two each by an identical light crystal, as shown in Fig.\
\ref{FIG-DNR-1}. In the absence of a which-way detector, {\sfD} and
{\sfE} are observed to interfere, as are {\sfF} and {\sfG}.
(Throughout this section, we shall adhere strictly to the notation of
\cite{DNR1998}.  \emph{Acknowledgement}: Fig.\ \ref{FIG-DNR-1} is a
simplified form of Fig.\ 1 of \cite{DNR1998}; Fig.\ \ref{FIG-DNR-2}
is a slightly modified form of Fig.\ 3 in the same source.)


The ground state $5^2S_{1/2}$ of $^{85}$Rb is split into two
hyperfine states with total angular momenta $F=2$ and $F=3$; denote
these two states by $|2\rangle$ and $|3\rangle$ respectively.  A
microwave field at $\omega_{\sf mw} \sim3$ GHz will induce Rabi
oscillations between the states $|2\rangle$ and $|3\rangle$. Let
$|e\rangle$ denote the $5^2P_{3/2}$ excited state of $^{85}$Rb. A
simplified level scheme of $^{85}$Rb is shown in Fig.\
\ref{FIG-DNR-2}(a). The incident beam {\sfA} is prepared in the state
$|2\rangle$.

The frequency $\omega_{\sf light}$ of the light crystal is tuned
halfway between the two transitions $|2\rangle\rightarrow|e\rangle$
and $|3\rangle\rightarrow|e\rangle$, as shown in Fig.\
\ref{FIG-DNR-2}. It is detuned from the two transitions by amounts
$\Delta_{2e}$ and $\Delta_{3e}$ that have the same absolute values
but opposite signs: $\Delta_{3e} > 0, \Delta_{2e} = -\Delta_{3e} <
0$. Therefore, as in classical optics, Bragg refraction from the
light crystal will shift the phase of a $|2\rangle$-beam by $\pi$,
but will leave the phase of a $|3\rangle$-beam unchanged. This phase
difference is used to induce a population difference between the
refracted and transmitted beams. 


Which-way information is stored \emph{in the beams themselves}, as
follows (see Fig.\ \ref{FIG-DNR-2}(b)). A $\pi/2$-microwave pulse at
$\omega_{\sf mw}$ is applied to the incident beam (prepared in the
state $|2\rangle$) before it reaches the first light crystal. This
pulse changes the incident beam {\sfA} to the superposition
$(|3\rangle + |2\rangle)/\sqrt2$.  The first light crystal splits
this beam and, additionally, changes the phase of the $|2\rangle$
component in the Bragg-refracted beam {\sfB} by $\pi$. The
transmitted beam is unaffected. A second $\pi/2$-microwave pulse now
changes the population of {\sfB} to $-|2\rangle$, and that of {\sfC}
to $|3\rangle$.

After the second microwave pulse has acted, the state vector of the
beam system is, up to normalization
\begin{equation}\label{DNR-EQ-2}
|\psi\rangle = |\psi_{\sfB}\rangle\otimes|2\rangle + 
               |\psi_{\sfC}\rangle\otimes|3\rangle.
\end{equation}
In the above, $|\psi_{\sfB,\sfC}\rangle$ describe only the state of
the centre of mass. The authors write that:
\begin{quote}

Equation (\ref{DNR-EQ-2}) shows that the internal state is correlated
with the way taken by the atom. The which-way information can be read
out later by performing a measurement of the internal atomic state.
The result of this measurement reveals which way the atom took: if
the internal state is found to be $|2\rangle$, the atom moved along
{\sfB}, otherwise along {\sfC}.

\end{quote}

After the second microwave pulse, the components {\sfB} and {\sfC}
are incident upon a second light crystal (Fig.\ \ref{FIG-DNR-1}).
This crystal splits each component, but does not affect their
populations. Thus {\sfD} and {\sfF} are in state $|2\rangle$ (the
phase is no longer critical), and {\sfF} and {\sfG} are in state
$|3\rangle$. The state vector is now proportional to

\begin{equation}\label{DNR-EQ-3}
|\psi\rangle = -|\psi_{\sfD}\rangle\otimes|2\rangle  
               +|\psi_{\sfE}\rangle\otimes|3\rangle
               +|\psi_{\sfF}\rangle\otimes|2\rangle  
               +|\psi_{\sfG}\rangle\otimes|3\rangle.
\end{equation}
One sees from Fig.\ \ref{FIG-DNR-1} that, in the interference region,
{\sfD} and {\sfE} overlap in space, as do {\sfF} and {\sfG}, but the
first pair has negligible overlap with the second pair.

However, (\ref{DNR-EQ-3}) predicts \emph{no interference} in either
pair, because $\langle2|3\rangle = 0$; and that is exactly what is
observed. We draw the reader's attention to a crucial difference
between the SEW equation (\ref{SCULLY-2}) and the DNR equation
(\ref{DNR-EQ-3}); the latter involves no quantity that is alien to
the atom.


\section{Galilei invariance and quantum mecha\-nics}
\label{SEC-MAIN}

At the root of our endeavour is the assumption that nonrelativistic
quantum mechanics has an invariance group, which is the inhomogeneous
Galilei group. In this section we shall spell out how this assumption
leads to an explanation of the phenomena observed by DNR.


\subsection{Particles of nonzero mass}
\label{SUBSEC-M-PARTICLES}

It was established by In\"on\"u and Wigner in 1952 that true unitary
irreducible representations of the Galilei group $G$ do not have a
particle interpretation \cite{IW1952}. The fact that free particles
in nonrelativistic quantum mechanics could be described by unitary
\emph{ray} representations of $G$ was established by Bargmann in 1954
\cite{VB1954}. There is a one-parameter family of such
representations, the parameter being the mass.  Equivalently, one
could say that the group $G$ has a one-parameter family of central
extensions $\tG_m$, and a particle of mass $m \neq 0$ corresponds to
a true representation of $\tG_m$. The parameter $m$ may be made
explicit in the group exponent: 
\begin{equation}\label{G-EXPONENT-2} 
\omega(g_1,g_2) = \rmi\[\exp\[m\hspace{.005em} \gamma(g_1,g_2).  
\end{equation}

The group $G$ is a ten-parameter group. Its generators are $H$ (time
translations), $\bmP$ (space translations), $\bmJ$ (rotations) and
$\bmK$ (boosts). The Casimir operators of this group are 
\begin{equation}\label{G-CENTRE} 
\bsm{P}^2\;\,\text{and}\;\,(\bsm{K}\times\bsm{P})^2.  
\end{equation} 
The group $\tG_m$ is an eleven-parameter group; it has an extra
generator, $I$, which commutes with every other generator and is
represented by the identity matrix in any irreducible unitary
representation. The commutator $[P_i, K_j] = 0$ in $G$ is replaced by
$[P_i, K_j] = \rmi\delta_{ij} m I$ in $\tG_m$, all other commutators
remaining the same. As a result, the Casimir operators of $\tG_m$
differ drastically from those of $G$. They are, in addition to $I$, 
\begin{eqnarray}  
U &=& H - \frac{1}{2m}\bmP^2,\label{CASIMIR-1}\\[4mm] 
\bmS^2 &=&\left(\bmJ-\frac1m\bmK\times\bmP\right)^2.\label{CASIMIR-2} 
\end{eqnarray} 
The spectrum of $\bmS^2$ is discrete; its eigenvalues are $s(s+1)$,
where $s$ is a nonnegative integer (or half-odd integer, for the
covering group).  That of $U$ is continuous, and fills the real line.
By analogy with thermodynamics, the spectral value $u$ of $U$ in an
irreducible representation is called the \emph{internal energy} of
the particle.  An irreducible unitary representation of $\tG_m$ is
characterized by the pair $(u, s)$.


\subsection{Internal energy and equivalence of representations}

In the following, we shall consider a fixed nonzero $m$, and shall
omit the subscript $m$ of $\tG_m$. We shall denote true
representations of $\tG$ by $\tD$, and ray representations of $G$ by
$D$ instead of $(D,\omega)$; the factor system will be displayed
through the mass $m$.

The standard parametrization of a group element $g$ of $G$ is
$$ g = (b, \bma, \bmv, R),$$
where $b$ is a time translation, $\bma$ a space translation, $\bmv$ a
pure Galilei transformation or boost and $R$ a rotation. With this
parametrization, we may write an element $\tg$ of $\tG$ as
$$ \tg = (\theta, b, \bma, \bmv, R),$$
where $\theta \in \bR$.  For simplicity,\footnote{Nonzero spins
present no difficulties; only the formulae become longer (see
\cite{LL1972}).} we shall restrict ourselves to representations
$\tD_{(u,s)}$ of $\tG$ with $s=0$. Let $\frH=L^2(\bR^3,\rmd\bmp)$ and
$\psi \in\frH$. The representation $\tD_{(u,0)}$, which we shall
write as $\tD_u$, is defined by
\begin{equation}\label{D_u}
\tD_u(\theta, b, \bma, \bmv, R)\psi(\bmp) = 
\rme^{\rmi[\theta+Eb+\bmp\cdot\bma]}\psi(R^{-1}(\bmp-m\bmv)),
\end{equation}
where $E$, the total energy, is a spectral value of $H$. Using
(\ref{CASIMIR-1}), the representation (\ref{D_u}) may be written 
as
$$
\tD_u(\theta, b, \bma, \bmv, R)\psi(\bmp) = 
\rme^{\rmi ub}\rme^{\rmi[\theta+ (\bmp^2/2m)b+\bmp\cdot\bma]}
\psi(R^{-1}(\bmp-m\bmv)),
$$
which may be written more compactly as
\begin{equation}\label{D_0}
\tD_u(\theta, g)\psi(\bmp) = \rme^{\rmi ub}\tD_0(\theta,g)\psi(\bmp).
\end{equation}
For any $u$, the operators $\{\tD(0,g)\}$ constitute a
ray representation $\{D_u(g)\}$ of $G$. Setting $\theta=0$ in
(\ref{D_0}), we find, by a slight change of notation, that
\begin{equation}\label{D-TRUE}
D_{\upr}(g) = \rme^{\rmi ub}D_{\upr-u}(g),
\end{equation}
where $\upr$ is arbitrary. Since the phase factor on the right
depends only on $g$, (\ref{D-TRUE}) establishes that $D_{\upr}$ and
$D_{\upr-u}$ are equivalent.  This mathematical equivalence
represents the physical fact that the zero of energy is arbitrary.

Consider now the direct sum $D_{u} \oplus D_{\upr}$.  Using
(\ref{D-TRUE}), we arrive at the formula 
\begin{equation}\label{D+D}
D_u(g)\oplus D_{\upr}(g) = \rme^{\rmi ub}[D_0(g)\oplus
D_{\upr-u}(g)].
\end{equation}
This shows that,\[\footnote{This fact, which has no parallel in the
theory of true representations, seems to have been first noticed by
L\'evy-Leblond in 1963 \cite{LL1963}.} although the representations
$D_u$ and $D_{\upr}$ are equivalent for all $u, \upr$, the
representation $D_u(g)\oplus D_{\upr}(g)$ is equivalent to
$D_0(g)\oplus D_{\upr-u}(g)$, and \emph{not} to $D_0(g)\oplus D_0(g)$
for $\upr\neq u$; although the zero of energy can be chosen
arbitrarily, the same zero has to be chosen for all energies, so that
energy differences remain physically meaningful.

Let $D_u, D_{\upr}$ be irreducible on $\frH_1, \frH_2$
respectively, $\frH = \frH_1\oplus\frH_2$ and $\Pi_1,
\Pi_2$ the projections $\Pi_1:\frH\rightarrow\frH_1$,
$\Pi_2: \frH\rightarrow\frH_2$. For $\psi\in\frH$, set
\begin{equation}\label{EQ-NOTATION}
\psi_1 = \Pi_1\psi, \;\,\psi_2 = \Pi_2\psi.
\end{equation}
Clearly,
\begin{equation}\label{EQ-IP}
(\psi_1,\psi_2) = 0.
\end{equation}
This orthogonality is due to the fact that the states $\psi_1$ and
$\psi_2$ belong to different subrepresentations of the direct
sum; \emph{it would continue to hold even when $u = \upr$}.


\subsection{Loss of interference in the DNR experiment}
\label{SUBSEC-LOSS}

The electronic configuration of an atom has a countable number of
discrete electronic energy levels, of which we shall only be
concerned with two or three at a time. All of these have the same
mass, say $m$, in nonrelativistic quantum mechanics. The spin of the
atom as a whole will play no explicit role in our considerations, and
therefore we shall set $s=0$ to reduce notational clutter. We now
introduce Galilei invariance into our considerations via the following
crucial assumption:

\begin{assump}\label{ASSUMP-1}
The states of an atom in a given energy level belong to an irreducible
ray representation $D_u$, with mass $m$, of the Galilei group $G$. 
{\rm (The zero of $u$ is arbitrary.)}
\end{assump}

In a single-particle interference experiment with a two-arm
interferometer, the geometry of the apparatus forces the incoming
wave function to split into the sum of two components. (This is to be
regarded as an empirical fact.) Suppose now that one of these
components is induced to make a transition to a different energy
level by interaction with a which-way detector.  Then the wave
function of the atom should become a superposition of two different
components with different energies, but essentially the same
momentum, and therefore the same kinetic energy. Most of the energy
difference has therefore to be attributed to the internal energy.
However, as the internal energy is constant in an irreducible
representation, this implies that interaction with the interferometer
with which-way detector \emph{splits the representation of the
incoming wave into a direct sum of two different irreducibles}, which
differ in their internal energies by the energy difference of the two
atomic levels.  We shall condense this into a proposition:

\begin{prop}\label{PROP-KEY}
Let the incoming particle be in a state $\psi_{\sf in} \in D_u$.  If
a which-way detector is present in one of the arms of the
interferometer, then the outgoing state of the particle, $\psi_{\sf
out}$, will belong to the representation $D_u\oplus D_{\upr}$,
$u\neq\upr$. In the notation of $(\ref{EQ-NOTATION})$, $\psi_{\sf
out}$ will be given by
\begin{equation}\label{EQ-KEY}
\psi_{\sf out} = \psi_1 + \psi_2,
\end{equation}
and the orthogonality relation $(\ref{EQ-IP})$ will hold. 
\end{prop}

In the DNR experiment (as in the SEW gedankenexperiment) the
which-way detector changes the energy state of the atom, and has
little effect on its wavelength. If assumption \ref{ASSUMP-1} is
valid, then proposition \ref{PROP-KEY} will be valid to a very good
approximation.

Equation (\ref{EQ-IP}) will explain the loss of interference. In
Feynman's ge\-dan\-ken\-experiment, loss of interference was a
statistical phenomenon due to the uncertainty principle; by contrast,
loss of interference due to the orthogonality  (\ref{EQ-IP}) may be
called a \emph{dynamical} phenomenon.


\section{Comparison of DNR and Galilei invariance explanations}
\label{SEC-DISCUSS}

The fundamental difference between the explanations offered by SEW,
on the one hand, and DNR and Galilei invariance, on the other hand is
the following. In the former, there is no interference because the
two states of the which-way detector -- which is physically distinct
from the atom -- are orthogonal; in the latter, there is no
interference because the \emph{two states of the atom} are
orthogonal. At first sight it might appear that the DNR explanation
is just the Galilei invariance explanation pared down to the
essentials: a moving two-level atom. The identifications $\psi_u =
\psi_B\otimes|2\rangle$ and $\psi_{\upr} = \psi_C\otimes|3\rangle$
turn (\ref{DNR-EQ-2}) and (\ref{EQ-KEY}) into each other.

However, the explanations offered by Galilei invariance and by DNR
are not equivalent. This inequivalence may be discussed using the
notion of \emph{quantum erasure} in atom interferometry.\footnote{The
notion of quantum erasure was introduced by Scully and Dr\"uhl in
1982 \cite{SD1982}. Walborn et al carried out an experiment with
photons in 2002 \cite{W2002}. Quantum erasure in atom interferometry
was discussed by SEW. Our remarks on quantum erasure apply only to
atom interferometry; Galilei invariance can say little that is useful
about photons.}

The term quantum erasure refers to the deletion of which-way
information from an atom. If an atom provides which-way information
by jumping from state $j$ to state $k$, this information may be
`erased' by forcing it to jump back from state $k$ to state $j$. If,
in the DNR scheme, atoms in state $|3\rangle$ are forced to jump back
into the state $|2\rangle$, the expression (\ref{DNR-EQ-2}) for
$|\psi\rangle$ will become 
$$|\psi\rangle = |\psi_{\sfB}\rangle\otimes|2\rangle + 
               |\psi_{\sfC}\rangle\otimes|2\rangle,   $$
i.e., interference will be restored. In the Galilei invariance
explanation, however, replacing $\psi_{\upr}$ by $\psi_u$ in
$\psi_{\sf out}$ given by (\ref{EQ-KEY}) will not change the
orthogonality condition $(\psi_1,\psi_2) = 0$. To restore
interference, a quantum eraser \emph{has to collapse the
representation $D_u\oplus D_{\upr}$ to a single irreducible $D$}. 

Whether or not this happens is testable in the laboratory, and is
discussed in Sec.\ \ref{SUBSEC-DNR-2}.


\section{Experimental tests}\label{SEC-EXPTS}

The two experiments suggested below may discriminate between the
various explanations of the results of which-way experiments in atom
interferometry.

\subsection{Own-goal experiment}\label{SUBSEC-OWN-GOAL}

The setup is the same as that shown in Fig.\ \ref{FIG-SEW}, with two
important differences:

\begin{enumerate}

\item There is only one cavity; cavity 2 is not present.

\item Cavity 1 is prepared in a coherent state, so that an atom
entering it will decay with probability 1.

\end{enumerate}

Condition 2 above means that which-way information cannot be obtained
in this experiment; the addition of extra photons should make no
detectable difference to a cavity prepared in a coherent state.  If
interference is not observed under this arrangement, it would support
the Galilei invariance argument, and throw doubt on the SEW version of
complementarity.  If, however, interference \emph{is} observed, then
it would decisively \emph{refute} the Galilei invariance argument.

\subsection{Modified DNR experiment}\label{SUBSEC-DNR-2}

Suppose that the own-goal experiment has been performed, and that its
result supports the Galilei invariance explanation. Then the
following experiment would be of considerable interest. 

Consider the beams {\sf D, E, F {\rm and} G} of Fig.\ \ref{FIG-DNR-1}.
After the two $\pi/2$-microwave pulses have acted as described (Sec.\
\ref{SEC-DNR}), {\sf D, F} will be in state $|2\rangle$, and {\sf E,
G} in state $|3\rangle$.

Let now a $\pi$-microwave pulse be applied to {\sf E}. It will change
the state of {\sf E} to $|2\rangle$. The experiment will consist of
looking for interference in the pair {\sf D, E}.  Then:\footnote{The
experiment could also be carried out with the beams {\sf G, F}. It
does not matter whether the state of {\sf E} is changed to
$|2\rangle$, or that of {\sf D} to $|3\rangle$.} 
\begin{enumerate}

\item If interference is \emph{not} observed, it would be additional
support for the Galilei invariance argument (as opposed to the
two-level atom argument): the direct sum of two copies of the same
representation \emph{has not collapsed} into a single irreducible
representation.

\item If interference \emph{is} observed, it would mean either
that (a)~the Galilei invariance argument is valid, but the two
component representations have collapsed into one; or that (b)~the
Galilei invariance argument is invalid. 

\end{enumerate}

In the opinion of the present author, observation of interference in
this experiment will have significant theoretical implications, but
it would be premature to speculate.

\vspace{2ex}\noindent {\bf Acknowledgement} The author would like to
thank Professor H Roos for his comments on an earlier version of this
article.



\section*{Appendix A: Ray representations}\label{SEC-RR}

\renewcommand{\theequation}{A.\arabic{equation}}

The following collection of definitions and formulae is tailored to
our needs. More detailed summaries may be found in \cite{LL1963},
\cite{LL1972} and \cite{SEN2010}. A complete mathematical account
will be found in Bargmann's original paper \cite{VB1954}.

Following Wigner and Bargmann, we shall denote operator rays by
boldface symbols: an operator ray is a collection $\bsm{A} =
\{\rme^{\rmi\alpha}A|\alpha\in\bR, A\; \text{fixed}\}$, where $A$ is
any operator on $\frH$. An operator $B\in\bsm{A}$ is a
\emph{representative} of the ray. Operator rays can be multiplied,
and a family of operator rays $\{\bsm{D}(g)|g\in G\}$ is said to
form a \emph{ray representation} of the group $G$ if
\begin{equation}\label{DEF-RR}
 \bsm{D}(g_1)\bsm{D}(g_2) = \bsm{D}(g_1g_2)\;\,\forall\;\,g_1, g_2\in
G.
\end{equation}
As rays cannot be added, it would be more convenient to work with
operators. Bargmann showed how this could be done without sacrificing
generality.

If $D(g), D(\gpr)$ are representatives of $\bsm{D}(g),\bsm{D}(\gpr)$
respectively, then one must have
\begin{equation}\label{RR-TO-TR}
D(g_1)D(g_2) = {\omega(g_1,g_2)}D(g_1g_2)
\end{equation}
where $\omega(g_1,g_2)$ is a complex number of modulus unity. It has
to satisfy the condition
\begin{equation}\label{FACTOR-COND-1}
\omega(g_1,g_2g_3)\omega(g_2,g_3) =
\omega(g_1,g_2)\omega(g_1g_2,g_3) 
\end{equation}

\noindent
which follows from the associativity of multiplication in $G$, and
the condition 
\begin{equation}\label{FACTOR-COND-2}
\omega(e,e) = 1
\end{equation}
which follows from $D(e) = I$. A continuous complex-valued function
of modulus unity on $G\times G$ that satisfies (\ref{FACTOR-COND-1})
and (\ref{FACTOR-COND-2}) is called a \emph{factor system} of $G$.
Equations (\ref{RR-TO-TR}), (\ref{FACTOR-COND-1}) and
(\ref{FACTOR-COND-2}) are the defining relations of ray
representations in terms of operators.  

Let $\varphi$ be a continuous complex-valued function of modulus
unity on $G$. If $D(g)$ is a representative of the ray $\bsm{D}(g)$,
then so is 
\begin{equation}\label{EQUIV-RR}
\Dpr(g)=\varphi(g)D(g).
\end{equation} 
If the $D(g)$ satisfy (\ref{RR-TO-TR}), then the $\Dpr(g)$ satisfy
\begin{equation}\label{RR-TO-TR-2}
\Dpr(g_1)\Dpr(g_2) = {\omegapr(g_1,g_2)}\Dpr(g_1g_2)
\end{equation}
where
\begin{equation}\label{EQUIV-FS}
{\omegapr(g_1,g_2)} = \displaystyle 
\frac{\varphi(g_1)\varphi(g_2)}{\varphi(g_1g_2)}\omega(g_1,g_2)\cdot
\end{equation}

\vspace{1mm}\noindent
Both (\ref{RR-TO-TR}) and (\ref{RR-TO-TR-2}) describe operator
multiplication in the same ray representation (\ref{DEF-RR}) of $G$.
We therefore define two factor systems $\omega(g_1,g_2)$ and
$\omegapr(g_1,g_2)$ to be \emph{equivalent} if there exists a
continuous complex-valued function $\varphi$ of modulus unity on $G$
such that (\ref{EQUIV-FS}) is satisfied. It is easily verified that
this is a true equivalence relation. It partitions the set of factor
systems of $G$ into equivalence classes. The main problem of the
theory of ray representations is to determine the set of equivalence
classes of factor systems of $G$. A factor system $\omega$ may be
written as
\begin{equation}\label{G-EXPONENT}
\omega(g, \gpr) = \[\exp\[\rmi\xi(g,\gpr),
\end{equation}
where $\xi$ is a real-valued function on $G\times G$, called a
\emph{group exponent}.

A ray representation may be specified by a set of operator
representatives and an equivalence class of factor systems. We shall
write the pair as $(D, \omega)$.

The factor system $\omega$ is said to be \emph{equivalent to unity}
if there exists a $\varphi$ such that $\omegapr = 1$. In this case
the ray representation is equivalent to a true representation.

\subsection{Unitary equivalence of ray representations}
\label{SUBSEC-UNI-EQ}

Let $G$ be a group that admits a nontrivial factor system, and let
$(D, \omega)$ and $(\Dpr, \omegapr)$ be two ray representations
of $G$:
\begin{eqnarray}
D(g_1)D(g_2)\! &=&\! {\omega(g_1,g_2)}D(g_1g_2)\label{EQ-ORD}\\[3mm]
\Dpr(g_1)\Dpr(g_2)&=& {\omegapr(g_1,g_2)}\Dpr(g_1g_2)\label{EQ-PRIME}
\end{eqnarray}

\noindent
The factor systems $\omega, \omegapr$, and therefore the
representations $(D, \omega),\; (\Dpr, \omegapr)$, are equivalent if
there exists a continuous function $\varphi(g)$ on $G$ such that
(\ref{EQUIV-FS}) is satisfied.

Next, let $U$ be a unitary operator and $\zeta(g)$ a
continuous real-valued function with $\zeta(e)= 0$ such that 
\begin{equation}\label{PROJ-EQUIV-1}
UD(g)U^{-1} = \rme^{\rmi\zeta(g)}\Dpr(g)\;\,\forall\;\,g\in G.
\end{equation}

\noindent
Multiplying (\ref{EQ-ORD}) by $U$ from the left and $U^{-1}$ on the
right, using (\ref{PROJ-EQUIV-1}), setting $\varphi(g) = \exp\, \rmi
\zeta(g)$ and using (\ref{EQUIV-FS}), we recover (\ref{EQ-PRIME}).
This shows that the unitary transformation $U$ transforms the
representation $(D, \omega)$ into the representation $(\Dpr,
\omegapr)$. The two ray representations are equivalent; their
operator representatives can be transformed into each other by a
unitary transformation, but the representative of a transformed
operator ray has to be chosen in accordance with (\ref{PROJ-EQUIV-1}).

The equivalence (\ref{PROJ-EQUIV-1}) has been called \emph{projective
equivalence} by L\'evy-Leblond \cite{LL1963}.


\section*{Appendix B: Further references}\label{SEC-FURTHER-REF}

\renewcommand{\theequation}{A.\arabic{equation}}

For a recent introduction to quantum optics, see the textbook by
Fox \cite{MF2006}. The monograph by Haroche and Raimond \cite{HR2006}
gives more detailed accounts of the topics they cover. 

The Nobel lectures of Chu \cite{SC1997} and Cohen-Tannoudji
\cite{CCT1997} provide very readable introductions the the
manipulation of atoms by light.

Jaeger, Shimony and Vaidman \cite{JSV1995}, and independently,
Englert \cite{BE1996} have obtained a duality relation between path
determination and fringe visibility in interferometry without using
any Robertson-Heisenberg inequality derived from a pair of
noncommuting self-adjoint operators. It was suggested, on this basis,
that ``the duality relation is logically independent of the
uncertainty relation'' \cite{BE1996}. This suggestion was
controverted by D\"urr and Rempe, who derived the duality relation
from the Robertson-Heisenberg uncertainty relation for a
suitably-chosen pair of self-adjoint operators \cite{DR2000}. It
should, however, be recalled that spectral theorems, going back to
Hilbert and von Neumann, ensure that for any subset $S$ of real
numbers, there exist an infinity of self-adjoint operators that have
$S$ as their common spectrum, so that the construction of D\"urr and
Rempe may be highly nonunique.  An introduction to spectral theory --
enough to justify the above statement -- may be found in Appendix~A6
of \cite{SEN2010}.

On the other hand, conditions other than the spectrum may inhibit
certain quantities, such as time in quantum mechanics and the phase
of a quantized electromagnetic field, from being described by
self-adjoint operators.  Recent reviews of the time-energy and the
number-phase uncertainty relations may be found in Busch
\cite{PB2008} and Busch et al \cite{BLPY2001} respectively.


\end{document}